%% file: stop_27sept.tex
\documentclass[aps,prl,twocolumn,showpacs,superscriptaddress,groupedaddress,amsmath,amssymb,floatfix]{revtex4}

\usepackage{graphicx}  
\usepackage{dcolumn}   
\usepackage{bm}        
\usepackage{amssymb}   

\def\met{\mbox{${\hbox{$E$\kern-0.6em\lower-.1ex\hbox{/}}}_T$}} 
\def\metVec{\mbox{${\hbox{$\vec{E}$\kern-0.6em\lower-.1ex\hbox{/}}}_T$}} 

\begin{document}

\centerline{\hspace*{8.3in} FERMILAB-PUB-10-393-E} 

\title{Search for pair production of the scalar top quark in the electron+muon final state }


\input author_list.tex

\begin{abstract}

We report the result of a search for the pair production of the lightest
supersymmetric
 partner of the top quark ($\tilde{t}_1$)
 in $p\bar{p}$ collisions at a center-of-mass energy of 1.96 TeV
 at the Fermilab Tevatron collider corresponding to an
 integrated luminosity of 5.4 fb$^{-1}$.
The scalar top quarks are assumed to decay into a $b$ quark, a charged
lepton, and a scalar neutrino ($\tilde{\nu}$), and
the search is performed in the electron plus muon final state. No
significant  excess of events above the standard model prediction is
detected, and  improved exclusion limits at the 95\% C.L. are 
set in the the
($M_{\tilde{t}_1}$,$M_{\tilde{\nu}}$) mass plane.

\end{abstract}


\maketitle

\newpage
  Supersymmetric theories  \cite{Fayet1977249} predict the existence of
scalar partners
for each of the standard model (SM) fermions.
 In the minimal supersymmetric standard model (MSSM) \cite{1981NuPhB.193..150D}, the
 mixing between the chiral states of the scalar partners of the
 SM fermions is greatest for the partners of the top quark  due to its
large Yukawa coupling \cite{Beenakker19983}. Thus, it is
 possible that the scalar top quark ($\tilde{t}_1$) is the
lightest squark and has the largest
production  cross section. If  $R$-parity \cite{Fayet1977249}
 is conserved, then scalar top quarks would be produced by $p\bar{p}$ collisions in pairs
 with the dominant  processes  being quark-antiquark annihilation and 
gluon fusion \cite{Beenakker19983}.

In this letter we report on a search for the production of $\tilde{t}_1 
\bar{\tilde{t}}_1$ pairs
in the $b\bar{b}e^{\pm}\mu^{\mp}\tilde{\nu}\overline{\tilde{\nu}}$ final
 state.
 We assume that the $\tilde{t}_1$ has
a 100\% branching fraction in this
three-body decay mode with equal fraction to each lepton type, that $R$-parity is
conserved, and that the sneutrino ($\tilde{\nu}$) is
the lightest supersymmetric particle or decays invisibly into a neutrino and a
neutralino ($\tilde{\chi}^0_{1}$). This analysis uses data corresponding to
an integrated luminosity of 5.4~fb$^{-1}$ collected
using the D0 detector operating at the Fermilab Tevatron collider at
$\sqrt{s}=1.96$~TeV. The data were collected from April 2002 through June 
2009.
 The D0 Collaboration has previously searched 
~\cite{PEDRAME,SQUARK_PUB,stop_runI}
for top squark pair production in the
final states $b\bar{b}\ell^{\pm}\ell^{\mp}\tilde{\nu}\overline{\tilde{\nu}}$
where the lepton pair is $ee$, $\mu\mu$, or $e\mu$. Two of these earlier 
searches used subsets of this data set corresponding to integrated
luminosities of 0.43~fb$^{-1}$ and 1.1~fb$^{-1}$, while the earliest 
search used data from the Tevatron Run I, corresponding to an integrated 
luminosity of $0.11$~fb$^{-1}$.
 Searches for top squark pair production in the
$b\bar{b}\ell^{\pm}\ell^{\mp}\tilde{\nu}\overline{\tilde{\nu}}$
final states have also been reported  by the CDF  collaboration 
\cite{CDF_RESULT} and by the ALEPH, L3, and OPAL Collaborations \cite{LEP}.

The main components of the D0 detector \cite{D0_det} include a central 
tracking system located
inside a 2~T superconducting solenoid. The inner-most tracking element
is the silicon microstrip tracker (SMT), followed by a scintillating
fiber tracker. These two detectors together measure the momenta of charged particles. The
tracking system provides full coverage in the azimuthal ($\phi$) direction for $|\eta|<2$,
 where the pseudorapidity $\eta$ is defined as $\eta=-\rm{ln}(\rm{tan}
\theta/2$) and $\theta$ is the polar angle with respect to the proton 
beam direction.
Outside the solenoid is the uranium/liquid argon calorimeter which is
divided into a central calorimeter and two end-cap calorimeters. Each of these
three calorimeters have electromagnetic layers followed by hadronic 
layers.
The outermost component of the detector is the muon system, which consists of proportional
drift tubes and scintillator trigger counters,
followed by 1.8~T iron toroids and two additional layers of  drift tubes 
and scintillators.
Events are selected for offline analysis by a three-level trigger system.
All events are required to pass one of a suite of single-electron 
triggers or single-muon triggers using information from the tracking 
system, the calorimeter, and the muon system. 

For each event, a primary $p\bar{p}$ interaction vertex is defined.
If more than one vertex is reconstructed, the primary  vertex is taken to 
be the vertex
least consistent with originating from a soft collision.
The location of the primary vertex along the beam direction
is required to be within $\pm$60 cm
of the detector center.

Jets are reconstructed using the D0 Run II cone algorithm
\cite{Blazey:2000qt}
  with cone of radius  $\mathcal{R}\equiv\sqrt{(\Delta \phi)^{2}~+~(\Delta 
y)^2}~<~0.5$,
  where $y$ is the rapidity.
  Jet energies are calibrated using the standard D0 procedure 
\cite{D0_JES}.
A jet is retained in an event if it has
transverse energy $E_T>$~20~GeV, $|\eta|~<~2.5$, and if
$\Delta\mathcal{R}(\text{jet,electron}) = \sqrt{(\Delta\phi)^{2} + 
(\Delta\eta)^{2}} > 0.5$.
 No requirement on the number of jets is applied.

Electrons are required to have transverse momentum $p^{e}_{T} > 15$ GeV and 
 $|\eta| < 1.1$. They are required to be isolated, defined as having
$[E_{tot}(0.4)-E_{EM}(0.2)]/E_{EM}(0.2)<0.15$ where
$E_{tot}(0.4)$ is the total calorimeter energy in a cone of radius
$\mathcal{R}=0.4$ and $E_{EM}(0.2)$ is the electromagnetic energy in 
a cone of radius $\mathcal{R}=0.2$.
In addition, the shower development in the calorimeter is required
to be consistent with that of an electromagnetic shower both  transversely and
longitudinally.
An eight-variable likelihood function is constructed to further distinguish
between electromagnetic and hadronic showers. The output of this function ranges from 0 to 1,
and electrons are required to have a likelihood value greater than 0.85.
 Electromagnetic showers associated with electrons are also required to match
a central track  within 
$\Delta\eta <0.05 $ and $\Delta\phi<0.05$ of the electromagnetic cluster.

Muons are required to have transverse momentum $p^{\mu}_{T} > 10$ GeV and $|\eta| < 2$.
They are required to have both drift tube and
scintillator hits in the muon system and to match a track in the central tracker.
If the central track includes hits in the SMT, the
distance of closest approach (DCA) between the muon track and the primary vertex is
required to be less  than 0.02 cm. If there are no SMT hits, then the DCA is
required to be less than 0.2~cm.
Muons are also required to satisfy isolation requirements in both the
calorimeter and the central tracker. For the calorimeter isolation,
the transverse energy in the cone $\mathcal{R} < 0.5$
around the muon track divided by $p^{\mu}_{T}$ must be less than 0.15.
For the central tracker isolation, the sum of the transverse energy
of the tracks in the hollow cone  $0.1<\mathcal{R}<0.5$
divided by $p^{\mu}_{T}$ must be less than 0.15.

Events are required to have exactly one electron and
one muon with opposite charge and to have a minimum separation between the electron and the muon
$\Delta\mathcal{R}(e,\mu) > 0.5$.
The missing transverse energy (\met) is calculated from the calorimeter energy
corrected for the jet and electron calibrations.
It is then adjusted to account for the transverse momentum of the muon.
All retained events are required to have \met $>$ 7 GeV. We refer to this 
preliminary set of
selection requirements as the preselection.

Signal Monte Carlo (MC) events are generated in a 2-D grid, i.e., for 
$\tilde{t}_1$ masses
ranging from 100~GeV to 240~GeV, and for $\tilde{\nu}$ masses ranging from 
40~GeV to 140~GeV, each in
10~GeV steps.
For each point, the MSSM decay
 parameters are calculated with {\sc suspect} \cite{Djouadi:2002ze}
and {\sc sdecay} \cite{sdecay}.
  {\sc madgraph/madevent}  \cite{Alwall:2007st}  is used to
generate four-vectors for the signal events with {\sc pythia} \cite{PYTHIA}
   providing  the showering and hadronization. The next-to-leading  order
(NLO) cross
section for $\tilde{t}_1$ pair production is calculated by {\sc prospino}
2.0
 \cite{PROSPINO} with the CTEQ6.1M~\cite{pumplin-2002-012} parton
distribution functions (PDFs).
The calculations are performed  with the factorization and renormalization scales
set to one, one half, and two times the $\tilde{t}_1$ mass to determine
the nominal value and the negative and positive uncertainties.
 The scale factor uncertainties are combined quadratically with the PDF
 uncertainties~\cite{pumplin-2002-012,stump-2003-0310}
to give the total theoretical uncertainties for the signal cross sections.

  The dominant SM backgrounds  for this decay are $Z/\gamma^{*} \to
\tau \tau$
with $\tau \to l \nu$; diboson production including $WW$, $WZ$, and $ZZ$; top
quark pairs; $W$ + jets; and  instrumental background coming from
multijet (MJ) processes where jets are misidentified
as electrons or contain muons that pass the isolation criterion
 and with \met\ arising from energy mismeasurement.
All the background processes in this analysis except for MJ are modeled
using MC simulation.
  Vector boson pair production is simulated  with {\sc pythia}, while all
other backgrounds are simulated at the parton level with {\sc
alpgen} \cite{Mangano:2002ea},  with {\sc pythia} used for hadronization
and showering. In order to simulate detector  noise and multiple 
$p\bar{p}$
interaction effects, each MC event is overlayed with a data event from
randomly chosen $p\bar{p}$ crossings.

MC correction factors determined from data are
applied to make distributions consistent between data and MC.
 These corrections include factors for the luminosity profile, beam spot
position, muon and electron identification efficiencies, boson transverse momentum,
and jet, electron, and muon energy resolutions.

 The MJ background is estimated from a selection of
  data events not overlapping with the search sample and is selected by 
inverting the
electron likelihood and muon isolation requirements.
This sample is used to determine the shape of the MJ
background. Because most same-sign di-lepton events come from MJ processes,
we obtain the normalization factor by taking the ratio
of the number of same-sign events that pass the likelihood and isolation
requirements to the number of
same-sign events that fail these requirements.
 To remove $W$+jet events from the MJ same-sign sample,
we make the additional requirement $\met <20$ GeV, since $W$+jets events tend to have
large \met.  We also correct this ratio for
non-MJ SM processes that produce like-sign leptons, using the MC samples.
 
 Data events are required to satisfy at least one of a suite of 
single-electron or single-muon triggers.
  The  efficiency of the combination of the single-electron triggers is 
measured using a
  subset of the search sample in which at least one of the single-muon 
triggers fired, and vice-versa for the single muon triggers. 
The combination of these two efficiencies, taken to be the overall trigger
efficiency,
 is then applied as a correction to the MC samples.
\begin {figure}[h]
\includegraphics[width=2.2in]{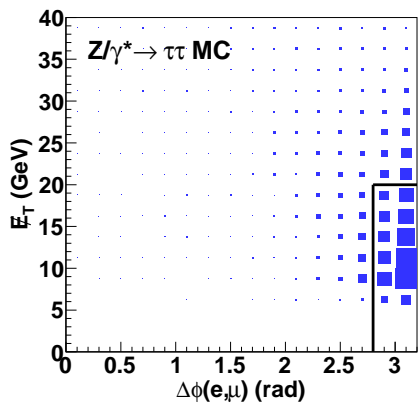}
\caption {(color online) \met\ versus  $\Delta \phi(e, \mu)$ for
$Z/\gamma^{*}\to\tau\tau$ MC events. Events inside the black box in the
lower right corner are removed. The sizes of the boxes are proportional
to the number of events in the underlying cell.}
\label{ztotau}
\end {figure}

The mass difference, $\Delta M=M_{\tilde{t}_1}-M_{\tilde{\nu}}$ determines
the kinematics of the final state.
A  larger $\Delta M$  will lead, on average, to larger \met, larger
jet energy,  and higher $p_T$ charged leptons. We divide the range of
$\Delta M$ into a ``large-$\Delta M$'' region ($\Delta M>60$ GeV) and 
``small-$\Delta M$'' region
($\Delta M < $60 GeV). To illustrate these regions, we have chosen two
benchmark points,
 ($M_{\tilde{t}_1},~M_{\tilde{\nu}}$)~=~(200~GeV,~100~GeV) and
 (110~GeV, 90~GeV), which will be referred to as the large-$\Delta M$
and small-$\Delta M$ benchmarks, respectively.
 Since there  are many signal points and their characteristics differ
significantly,
the analysis strategy is to optimize the signal selection as a function
of $\Delta M$.

For all values of $\Delta M$, the largest background
after preselection is $Z/\gamma^{*} \rightarrow\tau\tau$.
 A two-dimensional plot of the
azimuthal angle between the electron and muon, $\Delta \phi
(e,\mu)$, vs. \met\ for $Z/\gamma^{*}\to\tau\tau$ MC events is shown
in Fig. \ref{ztotau}.
The two leptons from $Z/\gamma^{*}\to\tau\tau$ tend to be
back-to-back in $\phi$,
and tend to have low
\met. We therefore reject events in which
$\Delta \phi(e,\mu) > 2.8$ and $\met < 20$ GeV and label this as
``Selection 1''.

\begin {figure}[h]
\includegraphics[width=2.6in]{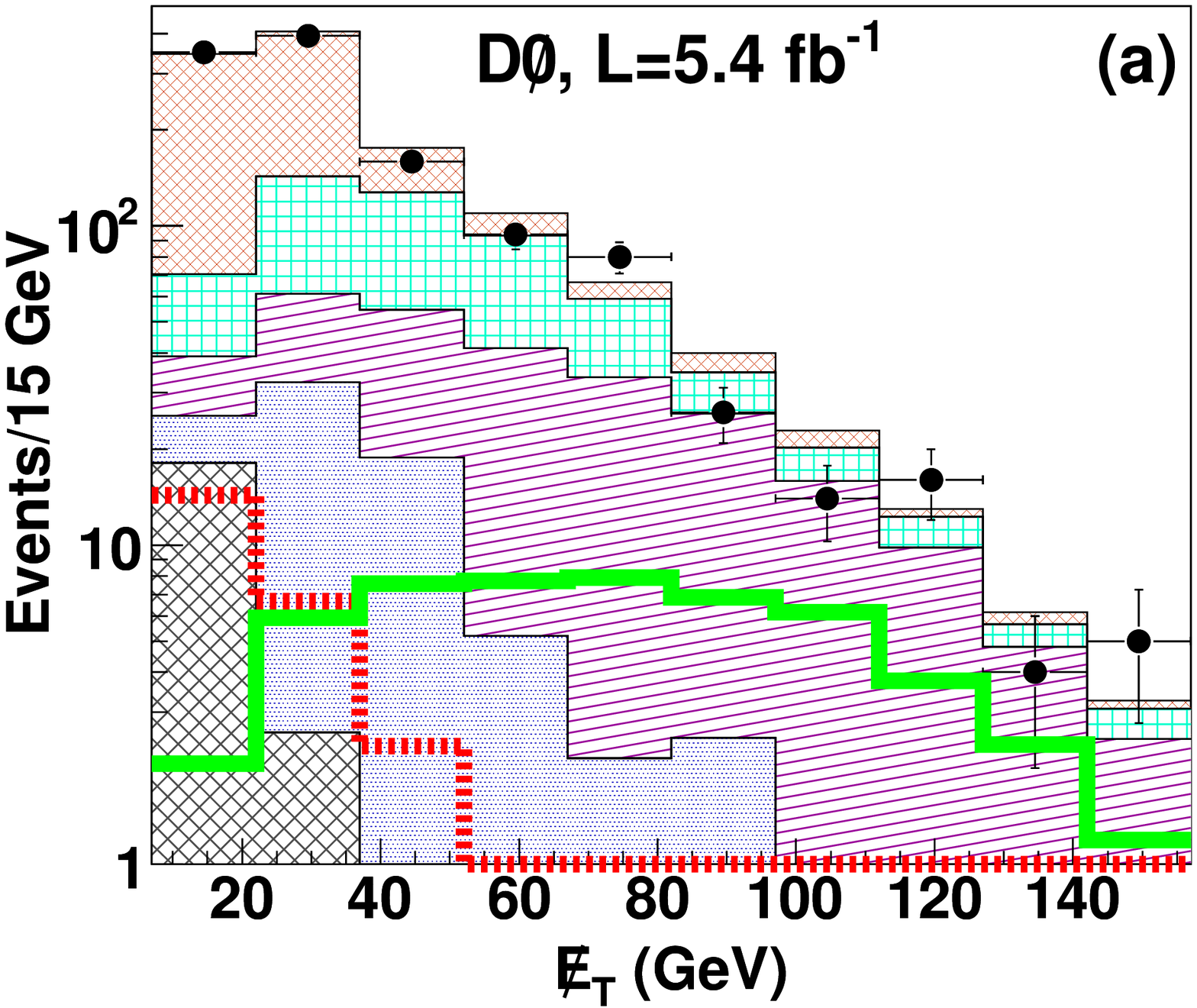}
\includegraphics[width=2.6in]{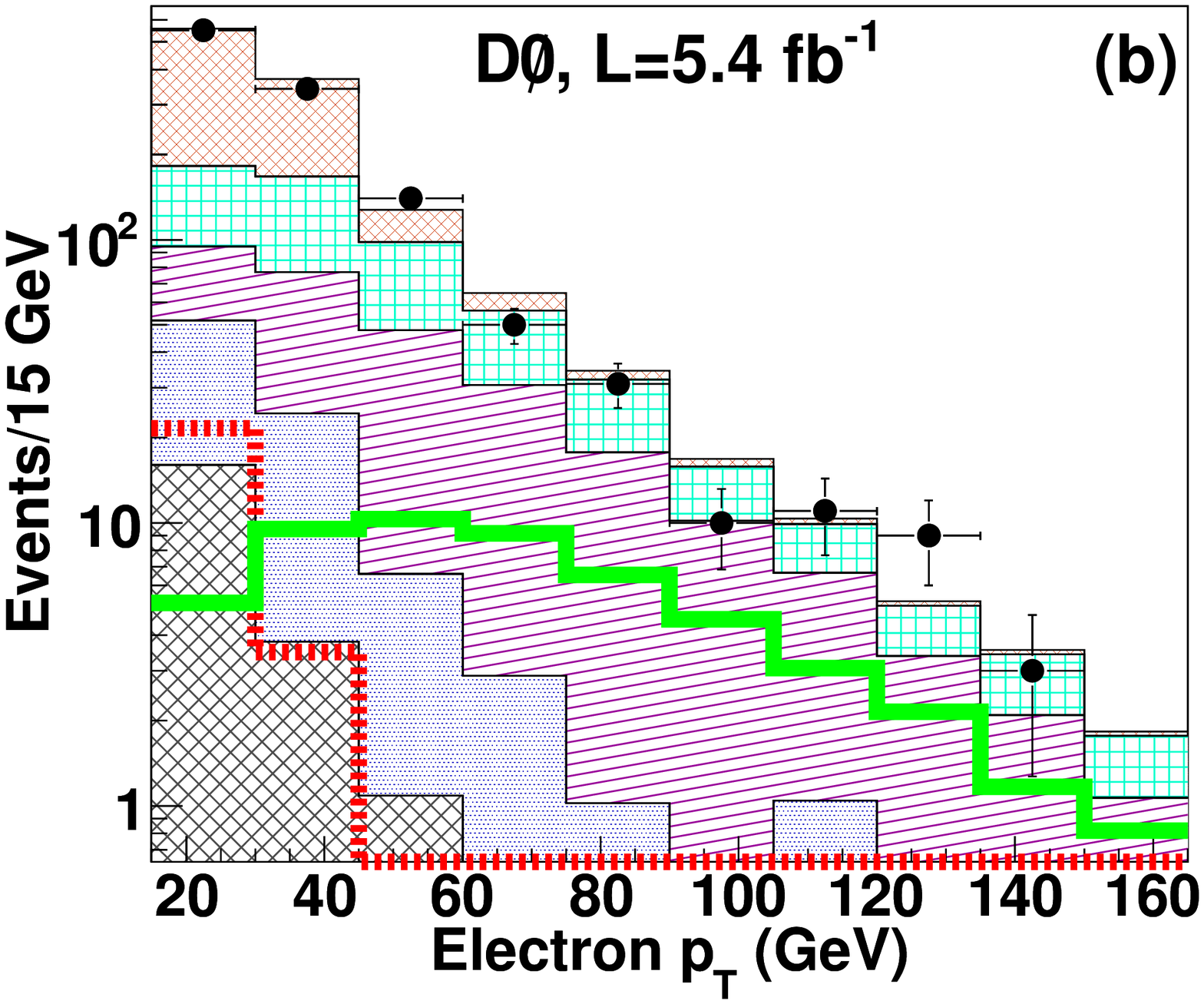}
\includegraphics[width=2.6in]{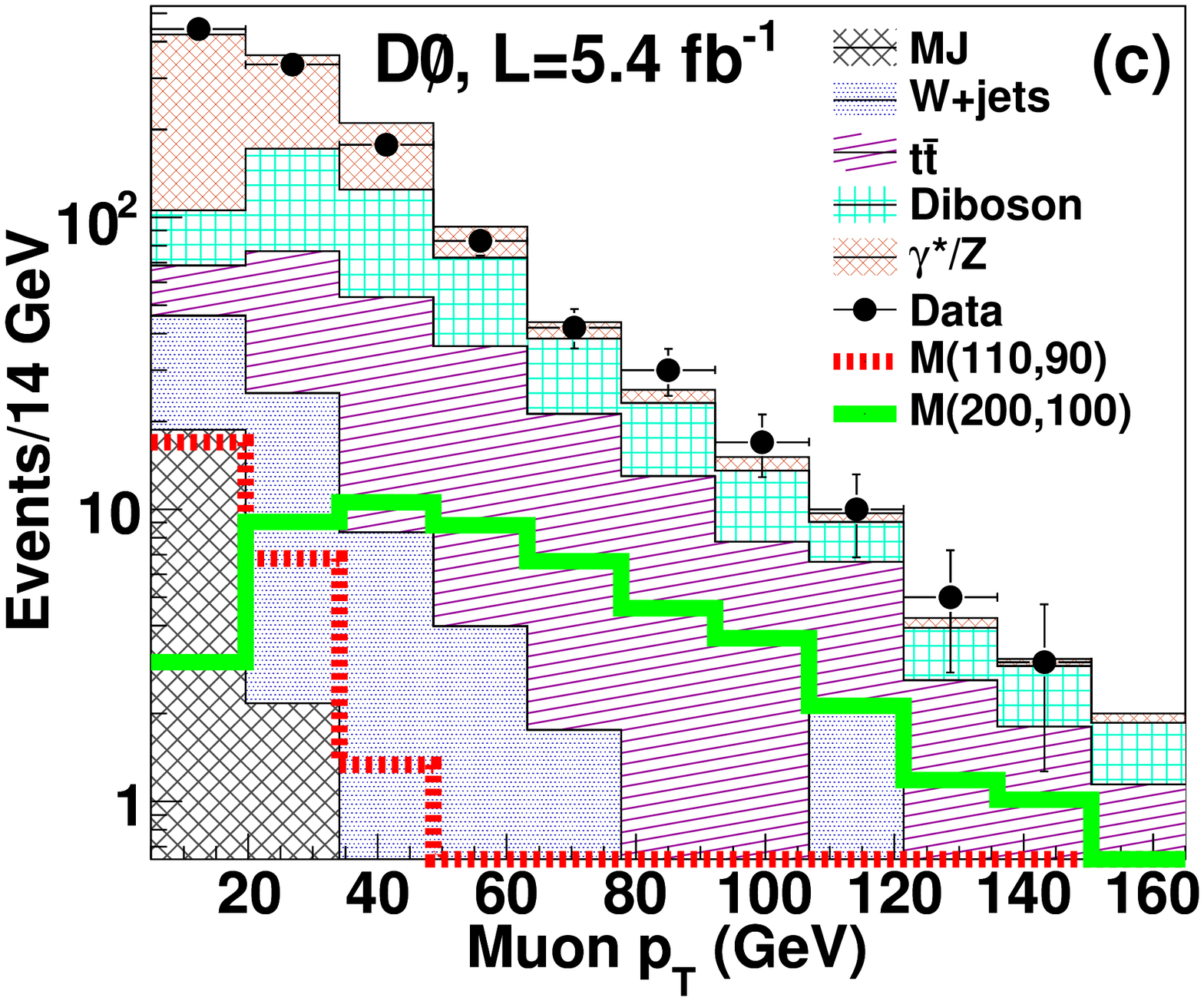}
\caption  {(color online) Distribution of (a) \met,
 (b) electron $p_T$, and (c) muon $p_T$ comparing data and all
background processes after Selection 1. 
The thick dashed and thick solid lines 
represent the small-$\Delta M$ and
large-$\Delta M$ signal benchmarks, respectively. }
\label {aftercut1}
\end {figure}

 Figure  \ref{aftercut1} compares \met, electron $p_T$,
 and muon $p_{T}$ of the data and the
sum of all backgrounds at this stage of the analysis. The agreement
confirms our understanding of the SM backgrounds,
of the trigger efficiency, and of other MC corrections.
After selection 1, the three largest backgrounds are
$Z/\gamma^{*}\rightarrow\tau\tau$,
$WW$, and $t\bar{t}$ production. To discriminate these backgrounds
from signal we create for each of them a composite
discriminant variable from a linear combination of
kinematic quantities.
We use the {\sc R} software package \cite{R_SOFTWARE} to calculate
the maximum  likelihood coefficients $\vec{\beta}$ for a generalized
linear model (GLM) \cite{NELDER} of the form
\begin{eqnarray}
\delta A = \ln{\frac{\mu}{1-\mu}} = \beta_0 + \vec{\beta} \cdot \vec{X}
\end{eqnarray}
to discriminate between signal and a specific background source $A$.
 Here, $\mu$ is the probability that an event is signal,
$\beta_0$ is a constant, $\vec{\beta}$ is
the vector of coefficients, and $\vec{X}$ is the vector of event
kinematic variables. By construction,
$\delta A=0$ when $\mu=0.5$, and signal-like events have positive $\delta
A$.
The discriminant $\delta Z$ is constructed to separate signal from
$Z/\gamma^* \to \tau \tau$ background, using an equal number of
signal and $Z/\gamma^* \to \tau \tau$ MC
events  to determine the coefficients $\beta_0$ and
$\vec{\beta}$. For $\vec{X}$ we use the following variables:
$\text{ln}(\met)$,
$\text{ln}(p_{T}^{\mu})$,
$\text{ln}(p_{T}^{e})$,
$\Delta\phi(e,\mu)$,
$\Delta\phi(e,\met)$,
$\Delta\phi(\mu,\met)$, and
$\Delta\phi(e,\met)\times\Delta\phi(\mu,\met)$.
For each value of $\Delta M$, ranging from 20 to 200 GeV,
we use the same variables with re-optimized coefficients.
We use a similar method for creating the
discriminants  $\delta WW$ and $\delta t\bar{t}$ to separate
 signal from  $WW$ and $t \bar{t}$ backgrounds.
For $\delta WW$ we use the variables
$\text{ln}(\met)$,
$\text{ln}(p_{T}^{\mu})$,
$\text{ln}(p_{T}^{e})$,
number of jets,
$\Delta\phi(e,\mu)$, and
$\text{ln} (WW_{\text{tag}})$.
Here $WW_{\text{tag}}$ is the magnitude of the vector sum
of $p_{T}^{e}$, $p_{T}^{\mu}$, and \met, which should be close to zero for 
 $WW$ events. 
For $\delta t\bar{t}$, we use the variables
$\text{ln}(\met)$,
$\text{ln}(p_{T}^{\mu})$,
$\text{ln}(p_{T}^{e})$,
$\text{ln}(1+H_{T})$,
the energy of the second most energetic jet, and
$WW_{\text{tag}}$. The variable $H_{T}$ is the scalar sum of the
transverse energies of all jets in an event.

\begin{figure}[th]
\begin{center}
\includegraphics[width=2.1in]{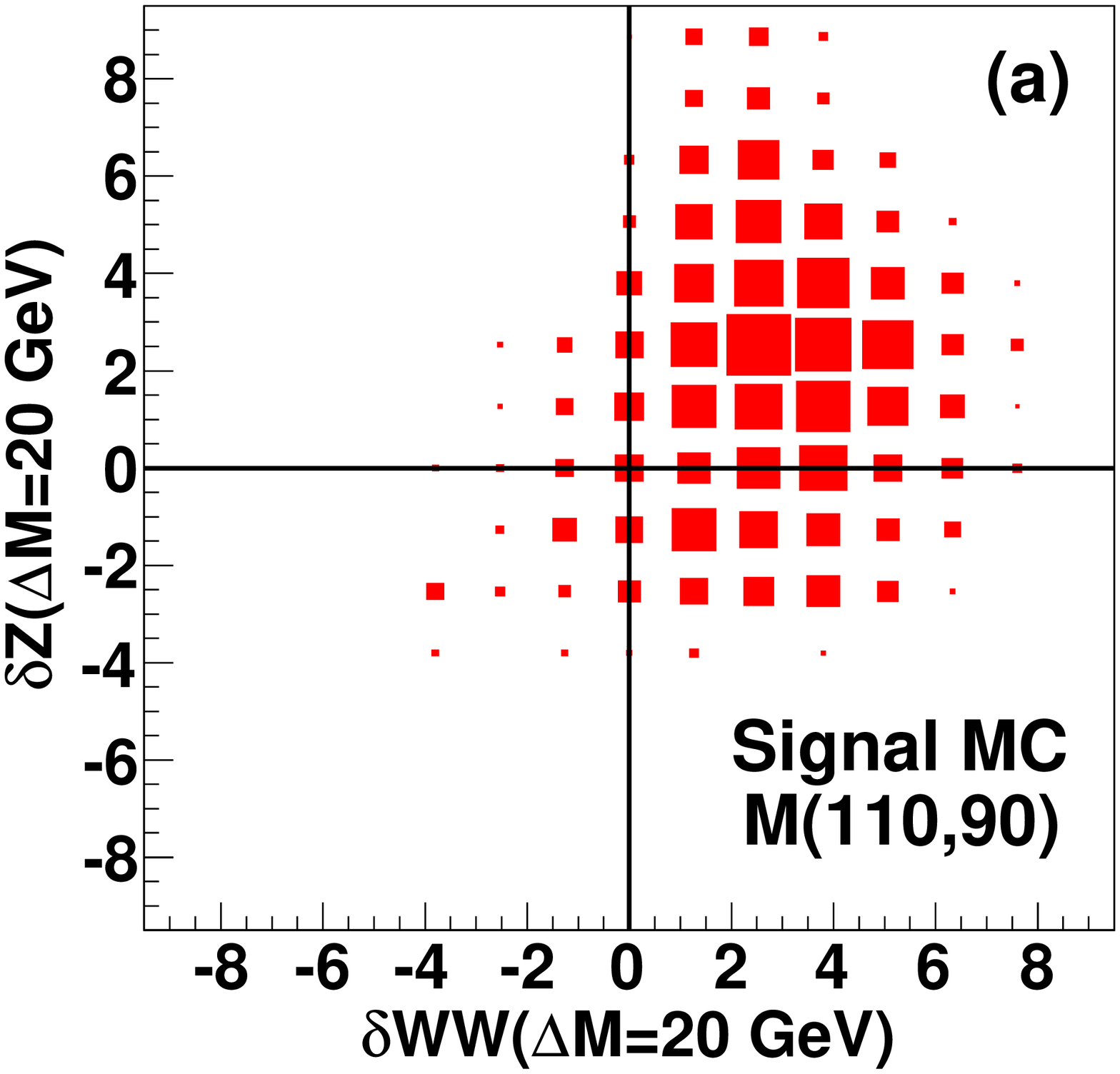}
\includegraphics[width=2.1in]{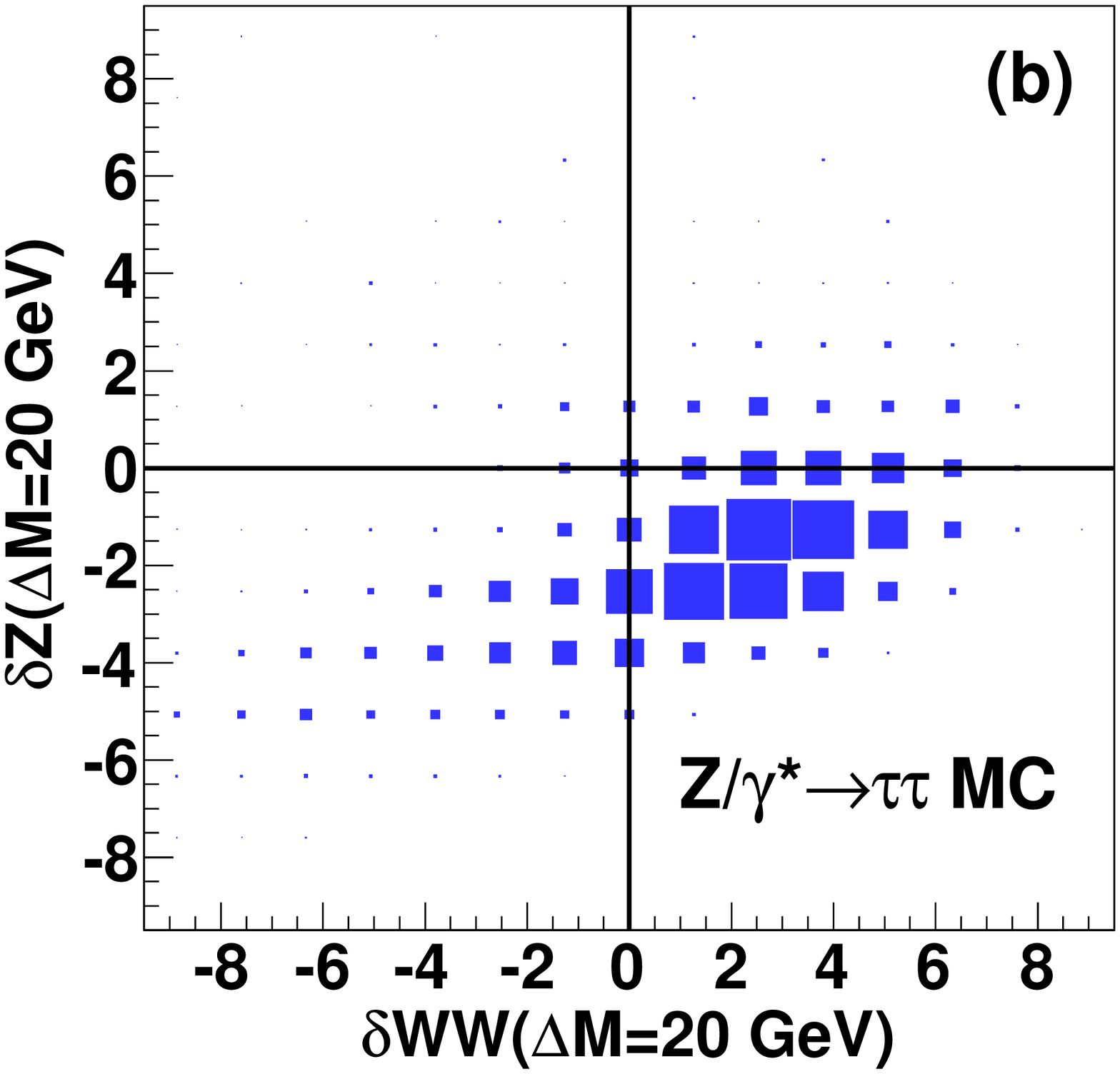}
\includegraphics[width=2.1in]{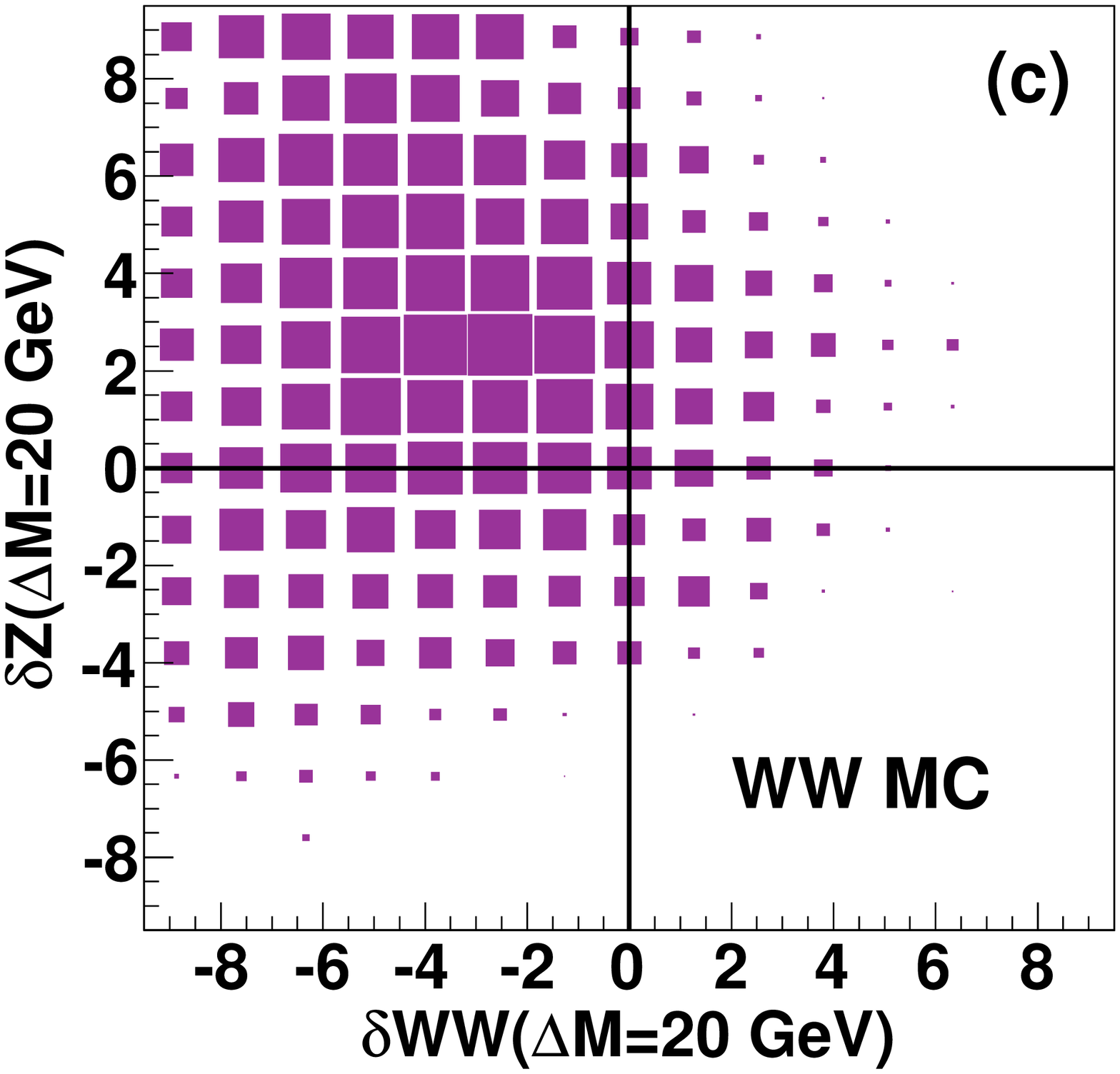}
\caption{(color online) Distribution of $\delta Z$ versus $\delta WW$ for
(a) the small-$\Delta M$ signal benchmark, ($M_{\tilde{t}_1}, 
M_{\tilde{\nu}}$) = (110 GeV, 90 GeV),  MC events,
(b) $Z/\gamma^{*} \to \tau \tau$ MC events, and
(c) $WW$ MC events.
The top right quadrant is used in the limit-setting procedure.  }
 \label{DISCRIMINANT}
\end{center}
\end{figure}

We first apply a requirement using the most effective discriminator of the
three. For $\Delta M < 60$ GeV, we require $\delta t\bar{t} > 0$.
The efficiency of this requirement is 0.95 for the small-$\Delta M$ signal benchmark
and 0.03 for $t\bar{t}$.
For $\Delta M \geq 60$ GeV, we require $\delta Z >0$.
The efficiency of this requirement is 0.96 for the large-$\Delta M$ signal benchmark
and 0.01 for $Z/\gamma^{*}\rightarrow\tau\tau$.
After making these requirements on one variable, we build 2-D
distributions of the two remaining discriminants.
Figure \ref{DISCRIMINANT} shows these distributions for the small-$\Delta M$ benchmark
signal and the two most significant remaining backgrounds.
In calculating the signal exclusion confidence limits, we use only the bins
in the upper right quadrant where the signal is concentrated.

Table \ref{summary} summarizes the expected backgrounds, the expected signal,
the observed data events, and the selection efficiencies.

\begin{table*}[ht]
\begin{center}

\begin{tabular}{r D{,}{\,\pm\,}{-1}  D{,}{\,\pm\,}{-1} D{,}{\,\pm\,}{-3}  D{,}{\,\pm\,}{-5} }
\hline\hline
 & \multicolumn{1}{c}{\hspace*{.2cm}Preselection\hspace*{.2cm}} & \multicolumn{1}{c}{Selection 1} & \multicolumn{2}{c}{Selection 2} \\[0.0 ex]

   \multicolumn{1}{c}{ }
 & \multicolumn{1}{c}{ }
 & \multicolumn{1}{c}{ }
 & \multicolumn{1}{c}{$\delta t \bar{t}>0$}
 & \multicolumn{1}{c}{$ \delta Z >0$}  \\[0 ex]

 Sample & \multicolumn{1}{c}{Events}
 & \multicolumn{1}{c}{Events}
 & \multicolumn{1}{c}{Events}
 & \multicolumn{1}{c}{Events}  \\[0 ex]
\hline
$Z\to \tau \tau$  &   1516 ,  150   &   582, 61    &   515 , 54   &  17.3 
, 2.0     \\ [1 ex]
$Z \to \mu \mu$   &   33.1 , 4.7    &   22.9 , 3.7    &   16.3 , 2.9    &  5.5 , 1.2   \\ [1 ex]
$Z \to ee$    &   23.2 , 3.9    &   16.6 , 3.0    &   10.2 , 2.2    &  0.1 , 2.3     \\ [1 ex]
$WZ$ &  12.7 , 1.6  &  12.0 , 1.5  &  6.3 , 0.8   &  9.3 , 1.2   \\ [1 ex]
$WW$    &   295 , 32   &   268 , 30   &   157 ,18    &  237 , 26    \\ [1 ex]
$ZZ$    &   2.2 , 0.3  &   2.0 , 0.3  &   1.0 , 0.15    &  1.1 , 0.16     \\ [1 ex]
$t\bar{t}$  & 206,28  &   204,28  &  6.6,0.9    & 179, 24 \\ [1 ex]
$W$  &   70 , 9.2   &   67.5, 9.0   &55,7.7& 53,7.4\\ [1 ex]
MJ    &   33 , 9.2   &   19.7 , 5.5   &18.4,5.1&1.3,0.35 \\ [0.0 ex]
\hline
\multicolumn{5}{c}{ } \\ [-1.95 ex]
Background total    &   2191,160 &1195, 73  &785,57&513,37\\ [0 ex]
Data  &  \multicolumn{1}{c}{2168}  &  \multicolumn{1}{c}{1147}  &  \multicolumn{1}{c}{776}   &  \multicolumn{1}{c}{472}  \\ [0.0 ex]
\hline
\multicolumn{1}{c}{Small-$\Delta M$ Benchmark}    &                                                
\\[0.0 ex]
\scriptsize{(110 GeV,90 GeV)}  &   35 , 5.6   &   25.5 , 4.2  &  23.8 , 3.9 &  \multicolumn{1}{c}{-}   \\ [1.0 ex]

\multicolumn{1}{c}{Large-$\Delta M$ Benchmark}    &                                                
\\[0.0 ex]
\scriptsize{(200 GeV,100 GeV)} &  55,9.3&53.4,9.0  &  \multicolumn{1}{c}{-} &51.8,8.7  \\ [0.5 ex]
\hline\hline
\end{tabular} \end{center}
 \caption{Expected numbers of background and signal events,
 and the number of events observed in the data at each stage of the analysis.
The errors include statistical and systematic uncertainties. 
For $\Delta M <$ 60~GeV,  Selection 2 is $\delta t \bar{t}(\Delta M=20$ 
~GeV)$ >0$, and for
$\Delta M\ge$~60 GeV, Selection 2
is $\delta Z(\Delta M=100$~GeV)$ >0$.  }
\label{summary}
\end{table*}

 The theoretical uncertainties on the signal cross section are 
approximately 20\% as discussed above
 and are the dominant uncertainties  in this analysis.
The uncertainty on the integrated luminosity is
6.1\%. Other systematic uncertainties included in the limit
setting calculations are
the lepton identification and track matching efficiencies (5\%),
the MJ background (27\%) scale factor,
 the jet energy calibration (1--2)\%,
and the production cross section uncertainties on
all the SM background processes (3--10)\%.
All uncertainties except for those on the MJ background and the SM production cross sections
are treated as fully correlated.
All systematic uncertainties are included in the limit calculations with 
Gaussian distributions \cite{collie}.

\begin{figure}[ht]
\begin{center}
\includegraphics[width=3.5in]{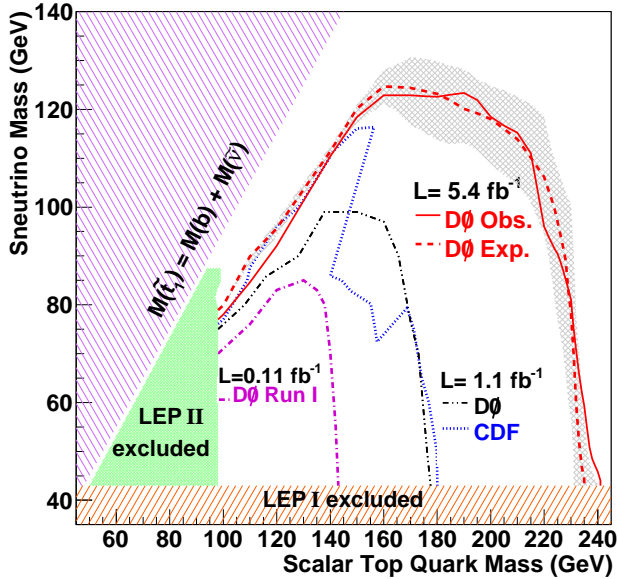}
  \caption{(color online) The observed (expected) 95\% C.L. exclusion
region  includes all mass points below the solid (dashed) line. The
 shaded band around the expected limit shows the effects of the scalar top
quark pair production cross section uncertainty.
  The kinematically forbidden region is represented in the upper left, and  
the
regions excluded by LEP~I and LEP~II \cite{LEP},
 by previous D0 searches \cite{SQUARK_PUB, stop_runI}, and by a previous 
CDF search \cite{CDF_RESULT} are also shown.  }
\label{EXCLUSION_PLANE}
\end{center}
\end{figure}

For $\Delta M < 60$ GeV,
we use the two dimensional histograms of the positive values of $\delta Z$
and
$\delta WW$ in the limit setting procedure. For $\Delta M \geq 60$ GeV,
we use the positive values of $\delta t\bar{t}$ and $\delta WW$.
A modified frequentist approach  \cite{Junk} is used to
determine the 95\% C.L. exclusion limits on scalar top quark production as a
function of the $\tilde{\nu}$  and $\tilde{t_1}$ masses, as shown in Fig.
\ref{EXCLUSION_PLANE}.
   Also shown are the exclusion regions from the CERN LEP 
experiments \cite{LEP},
   previous D0 searches \cite{SQUARK_PUB,stop_runI}, and a CDF search
\cite{CDF_RESULT}.

In conclusion,
 we set 95\% C.L. exclusion limits on the cross section
for scalar top quark pair production assuming a 100\% branching
fraction to
$b\bar{b}l^{\pm}l^{\mp}\tilde{\nu}\bar{\tilde{\nu}}$ using 5.4 fb$^{-1}$
of integrated luminosity from the D0 experiment at the Fermilab Tevatron 
collider.
We have excluded stop pair production for
$M_{\tilde{t}_1} < 210$ GeV when $M_{\tilde{\nu}} < 110$ GeV and the 
difference
$M_{\tilde{t}_1} - M_{\tilde{\nu}} > 30$ GeV. This extends the previous
limits on the top squark mass by more than 40 GeV for sneutrino masses less
than 90 GeV and the limits on the sneutrino mass by more than 30 GeV for top squark
mass equal to 150 GeV.

\input acknowledgement.tex
\bibliographystyle{apsrev}

\bibliography{stop_pub}

%
%
%
%
%
%
%
%
%
%
%
%
\end{document}

%% file: author_list.tex
\affiliation{Universidad de Buenos Aires, Buenos Aires, Argentina}
\affiliation{LAFEX, Centro Brasileiro de Pesquisas F{\'\i}sicas, Rio de Janeiro, Brazil}
\affiliation{Universidade do Estado do Rio de Janeiro, Rio de Janeiro, Brazil}
\affiliation{Universidade Federal do ABC, Santo Andr\'e, Brazil}
\affiliation{Instituto de F\'{\i}sica Te\'orica, Universidade Estadual Paulista, S\~ao Paulo, Brazil}
\affiliation{Simon Fraser University, Vancouver, British Columbia, and York University, Toronto, Ontario, Canada}
\affiliation{University of Science and Technology of China, Hefei, People's Republic of China}
\affiliation{Universidad de los Andes, Bogot\'{a}, Colombia}
\affiliation{Charles University, Faculty of Mathematics and Physics, Center for Particle Physics, Prague, Czech Republic}
\affiliation{Czech Technical University in Prague, Prague, Czech Republic}
\affiliation{Center for Particle Physics, Institute of Physics, Academy of Sciences of the Czech Republic, Prague, Czech Republic}
\affiliation{Universidad San Francisco de Quito, Quito, Ecuador}
\affiliation{LPC, Universit\'e Blaise Pascal, CNRS/IN2P3, Clermont, France}
\affiliation{LPSC, Universit\'e Joseph Fourier Grenoble 1, CNRS/IN2P3, Institut National Polytechnique de Grenoble, Grenoble, France}
\affiliation{CPPM, Aix-Marseille Universit\'e, CNRS/IN2P3, Marseille, France}
\affiliation{LAL, Universit\'e Paris-Sud, CNRS/IN2P3, Orsay, France}
\affiliation{LPNHE, Universit\'es Paris VI and VII, CNRS/IN2P3, Paris, France}
\affiliation{CEA, Irfu, SPP, Saclay, France}
\affiliation{IPHC, Universit\'e de Strasbourg, CNRS/IN2P3, Strasbourg, France}
\affiliation{IPNL, Universit\'e Lyon 1, CNRS/IN2P3, Villeurbanne, France and Universit\'e de Lyon, Lyon, France}
\affiliation{III. Physikalisches Institut A, RWTH Aachen University, Aachen, Germany}
\affiliation{Physikalisches Institut, Universit{\"a}t Freiburg, Freiburg, Germany}
\affiliation{II. Physikalisches Institut, Georg-August-Universit{\"a}t G\"ottingen, G\"ottingen, Germany}
\affiliation{Institut f{\"u}r Physik, Universit{\"a}t Mainz, Mainz, Germany}
\affiliation{Ludwig-Maximilians-Universit{\"a}t M{\"u}nchen, M{\"u}nchen, Germany}
\affiliation{Fachbereich Physik, Bergische  Universit{\"a}t Wuppertal, Wuppertal, Germany}
\affiliation{Panjab University, Chandigarh, India}
\affiliation{Delhi University, Delhi, India}
\affiliation{Tata Institute of Fundamental Research, Mumbai, India}
\affiliation{University College Dublin, Dublin, Ireland}
\affiliation{Korea Detector Laboratory, Korea University, Seoul, Korea}
\affiliation{CINVESTAV, Mexico City, Mexico}
\affiliation{FOM-Institute NIKHEF and University of Amsterdam/NIKHEF, Amsterdam, The Netherlands}
\affiliation{Radboud University Nijmegen/NIKHEF, Nijmegen, The Netherlands}
\affiliation{Joint Institute for Nuclear Research, Dubna, Russia}
\affiliation{Institute for Theoretical and Experimental Physics, Moscow, Russia}
\affiliation{Moscow State University, Moscow, Russia}
\affiliation{Institute for High Energy Physics, Protvino, Russia}
\affiliation{Petersburg Nuclear Physics Institute, St. Petersburg, Russia}
\affiliation{Stockholm University, Stockholm and Uppsala University, Uppsala, Sweden }
\affiliation{Lancaster University, Lancaster LA1 4YB, United Kingdom}
\affiliation{Imperial College London, London SW7 2AZ, United Kingdom}
\affiliation{The University of Manchester, Manchester M13 9PL, United Kingdom}
\affiliation{University of Arizona, Tucson, Arizona 85721, USA}
\affiliation{University of California Riverside, Riverside, California 92521, USA}
\affiliation{Florida State University, Tallahassee, Florida 32306, USA}
\affiliation{Fermi National Accelerator Laboratory, Batavia, Illinois 60510, USA}
\affiliation{University of Illinois at Chicago, Chicago, Illinois 60607, USA}
\affiliation{Northern Illinois University, DeKalb, Illinois 60115, USA}
\affiliation{Northwestern University, Evanston, Illinois 60208, USA}
\affiliation{Indiana University, Bloomington, Indiana 47405, USA}
\affiliation{Purdue University Calumet, Hammond, Indiana 46323, USA}
\affiliation{University of Notre Dame, Notre Dame, Indiana 46556, USA}
\affiliation{Iowa State University, Ames, Iowa 50011, USA}
\affiliation{University of Kansas, Lawrence, Kansas 66045, USA}
\affiliation{Kansas State University, Manhattan, Kansas 66506, USA}
\affiliation{Louisiana Tech University, Ruston, Louisiana 71272, USA}
\affiliation{University of Maryland, College Park, Maryland 20742, USA}
\affiliation{Boston University, Boston, Massachusetts 02215, USA}
\affiliation{Northeastern University, Boston, Massachusetts 02115, USA}
\affiliation{University of Michigan, Ann Arbor, Michigan 48109, USA}
\affiliation{Michigan State University, East Lansing, Michigan 48824, USA}
\affiliation{University of Mississippi, University, Mississippi 38677, USA}
\affiliation{University of Nebraska, Lincoln, Nebraska 68588, USA}
\affiliation{Rutgers University, Piscataway, New Jersey 08855, USA}
\affiliation{Princeton University, Princeton, New Jersey 08544, USA}
\affiliation{State University of New York, Buffalo, New York 14260, USA}
\affiliation{Columbia University, New York, New York 10027, USA}
\affiliation{University of Rochester, Rochester, New York 14627, USA}
\affiliation{State University of New York, Stony Brook, New York 11794, USA}
\affiliation{Brookhaven National Laboratory, Upton, New York 11973, USA}
\affiliation{Langston University, Langston, Oklahoma 73050, USA}
\affiliation{University of Oklahoma, Norman, Oklahoma 73019, USA}
\affiliation{Oklahoma State University, Stillwater, Oklahoma 74078, USA}
\affiliation{Brown University, Providence, Rhode Island 02912, USA}
\affiliation{University of Texas, Arlington, Texas 76019, USA}
\affiliation{Southern Methodist University, Dallas, Texas 75275, USA}
\affiliation{Rice University, Houston, Texas 77005, USA}
\affiliation{University of Virginia, Charlottesville, Virginia 22901, USA}
\affiliation{University of Washington, Seattle, Washington 98195, USA}
\author{V.M.~Abazov} \affiliation{Joint Institute for Nuclear Research, Dubna, Russia}
\author{B.~Abbott} \affiliation{University of Oklahoma, Norman, Oklahoma 73019, USA}
\author{M.~Abolins} \affiliation{Michigan State University, East Lansing, Michigan 48824, USA}
\author{B.S.~Acharya} \affiliation{Tata Institute of Fundamental Research, Mumbai, India}
\author{M.~Adams} \affiliation{University of Illinois at Chicago, Chicago, Illinois 60607, USA}
\author{T.~Adams} \affiliation{Florida State University, Tallahassee, Florida 32306, USA}
\author{G.D.~Alexeev} \affiliation{Joint Institute for Nuclear Research, Dubna, Russia}
\author{G.~Alkhazov} \affiliation{Petersburg Nuclear Physics Institute, St. Petersburg, Russia}
\author{A.~Alton$^{a}$} \affiliation{University of Michigan, Ann Arbor, Michigan 48109, USA}
\author{G.~Alverson} \affiliation{Northeastern University, Boston, Massachusetts 02115, USA}
% ...
%
%
\collaboration{The D0 Collaboration\footnote{with visitors from
$^{a}$Augustana College, Sioux Falls, SD, USA,
$^{b}$The University of Liverpool, Liverpool, UK,
$^{c}$SLAC, Menlo Park, CA, USA,
$^{d}$ICREA/IFAE, Barcelona, Spain,
$^{e}$Centro de Investigacion en Computacion - IPN, Mexico City, Mexico,
$^{f}$ECFM, Universidad Autonoma de Sinaloa, Culiac\'an, Mexico,
and 
$^{g}$Universit{\"a}t Bern, Bern, Switzerland.%
}} \noaffiliation
\vskip 0.25cm

%% file: acknowledgement.tex
%
We thank the staffs at Fermilab and collaborating institutions,
and acknowledge support from the
DOE and NSF (USA);
CEA and CNRS/IN2P3 (France);
FASI, Rosatom and RFBR (Russia);
CNPq, FAPERJ, FAPESP and FUNDUNESP (Brazil);
DAE and DST (India);
Colciencias (Colombia);
CONACyT (Mexico);
KRF and KOSEF (Korea);
CONICET and UBACyT (Argentina);
FOM (The Netherlands);
STFC and the Royal Society (United Kingdom);
MSMT and GACR (Czech Republic);
CRC Program and NSERC (Canada);
BMBF and DFG (Germany);
SFI (Ireland);
The Swedish Research Council (Sweden);
and
CAS and CNSF (China).